# Observation of multi Dirac fermion cloning induced by moiré potential in graphene-SiC heterostructure

C. L. Wu[1§], Q. Wan[1§], C. Peng[1§], S. K. Mo[1], R. Z. Li[1], K. M. Zhao[1], Y. P. Guo[2], C. D. Zhang[2] and N. Xu[1,*]

[1] *The Institute of Advanced Studies, Wuhan University, Wuhan 430072, China*
[2] *School of Physics and Technology, Wuhan University, Wuhan 430072, China*

§ There authors contributed equally.
* E-mail: nxu@whu.edu.cn

We reexamine the electronic structure of graphene on SiC substrate by angle-resolved photoemission spectroscopy. We directly observed multiply cloning of Dirac cone, in addition to ones previously attributed to $6\sqrt{3} \times 6\sqrt{3}R30°$ reconstruction. The locations, relative distances and anisotropy of Dirac cone replicas fully agree with the moiré pattern of graphene-SiC heterostructure. Our results provide a straightforward example of moiré potential modulation in engineering electronic structure with Dirac fermions.

Designing Dirac cone structure of graphene by Moiré heterostructures has become a promising platform for realizations and investigations of novel quantum phases and phase transitions [1-36]. In graphene-boron nitride (BN) heterostructure, moiré potential generates clones of Dirac cones at the corners of the moiré Brillouin zone (BZ), which leads to a fractal energy spectrum in a strong magnetic field [1-7]. Recently, moiré potential in twisted bilayer graphene [8-9] generate nearly flat bands with strong electronic correlation, and correlated insulator, superconductivity and exotic quantum phases are achieved [10-24]. Introducing additional superlattice potential from BN or WSe$_2$ into the twisted bilayer graphene and trilayer graphene, time reversal symmetry is breaking with ferromagnetism, superconductivity and quantum anomalous Hall effect realized [25-36].

Here we investigate the electronic structure of graphene-SiC heterostructure by angle-resolved photoemission spectroscopy (ARPES). Besides ones previously attributed to $6\sqrt{3} \times 6\sqrt{3}R30°$ reconstruction, we directly observed multiply Dirac cone replica in the first Brillouin zones of graphene. We ascertain that the location, relative distances and Fermi velocity anisotropy are fully consistent with the conner of the moiré Brillouin zones ($K_m$) of graphene-SiC heterostructure, for all the observed Dirac cone cloning. The moiré potential induce Van Hove singularities between the neighbors of Dirac cone clones, but well below $E_F$. Our results provide a straightforward example of moiré potential in engineering electronic structure with Dirac fermions.

High-quality, large-scale monolayer graphene was achieved by annealing process [37] of n-type 6H SiC(0001) purchased from PrMat. ARPES measurements were performed at the home-designed facility with a vacuum better than $5 \times 10^{-11}$ Torr and at a temperature lower than 6 K, with photon energy hv = 21.2 eV. Samples were directly transferred into the ARPES chamber, without breaking the vacuum.

The structure of graphene-SiC heterostructure is shown in Fig. 1a, with the primitive vectors appended. Figure 1b displays the lattices and primitive vectors of graphene (labeled as $\vec{G}_{Gra}(0,1)/\vec{G}_{Gra}(1,0)$) and SiC (labeled as $\vec{G}_{SiC}(0,1)/\vec{G}_{SiC}(1,0)$) in the reciprocal space. The corresponding vectors from the Γ to K points are also indexed as $\vec{K}_{Gra}(0,1)/\vec{K}_{Gra}(1,0)$ and $\vec{K}_{SiC}(0,1)/\vec{K}_{SiC}(1,0)$ for graphene and SiC, respectively.

Figure 2a shows Fermi surface (FS) map in the $k_x$-$k_y$ plane. Besides the intensity at $K_{Gra}(1,0)$ point corresponding to the original Dirac cone of graphene, additional hot spots clearly appears within the first BZ of graphene in the ARPES result. As seen from Fig. 2b-f, the band dispersions cutting through these hot spots are Dirac cone-like, similar as the main Dirac cone at the K point of graphene (Fig. 2g). We label all the hot spots by colored dots in Fig. 2a, based on the experimentally observed ones and symmetries. We note that the relative distance between the hot spot in cut #2 (Fig. 2c) label as purple dot and the one in cut #4 (Fig. 2d) but on the other side (light green dot) shows a good agreement with the length of $\vec{K}_{Gra}(1,1)$ ~ 1.7 Å. Furthermore, The relative distance between the hot spot in cut #3 and previously attributed to $6\sqrt{3} \times 6\sqrt{3}R30°$ reconstruction on the other side (hot spot as labeled by dark green dot, which has more stronger spectra weight in ARPES experiment using higher energy photons as seen from Supplementary Materials) fits the length of $\vec{G}_{SiC}(1,0)$ ~ 2.363 Å (twice of Γ-M in SiC). Later we will demonstrate that these relationships hold in an exact way.

As seen from Fig. 2h, the constant energy plot away from Dirac point shows a triangle-like shape at the K$_{Gra}$ point, instead of a circle. It is possibly due to the Fermi velocity anisotropy of Dirac cone in graphene along the Γ-K and K-M directions. Interestingly, for the Dirac cones shown in #2 and #4 cuts, the triangle-like shapes in the constant energy plot at high binding energy show an opposite direction to the one origin from graphene at the K point (Fig. 2h).

The location and anisotropy of the experimentally observed Dirac cone replicas are fully consistent with the moiré structures formed by graphene-SiC heterostructure. Along the $\vec{G}_{Gra}(1,1)$ direction (the k$_y$ direction) in Fig. 3a, the G$_{SiC}$(2,1) point sits between the G$_{Gra}$(1,1) and G$_{Gra}$(2,2) points. According to the convolution theorem of Fourier transformation, two types of hexagonal moiré structures are formed with primitive vectors of moiré BZ corresponding to
$$\vec{G}_{m1} = \vec{G}_{Gra}(1,1) - \vec{G}_{SiC}(2,1) \text{ and } \vec{G}_{m2} = \vec{G}_{Gra}(2,2) - \vec{G}_{SiC}(2,1).$$
The corresponding Γ-K vectors of these moiré structures are
$$\vec{K}_{m1} = \vec{K}_{Gra}(1,1) - \vec{K}_{SiC}(2,1) \text{ and } \vec{K}_{m2} = \vec{K}_{Gra}(2,2) - \vec{K}_{SiC}(2,1).$$
Note that
$$\vec{K}_{SiC}(2,1) = \vec{G}_{SiC}(1,0).$$
We direct append the $\vec{K}_{m1}$ and $\vec{K}_{m2}$ vectors to the experimental determined FS map of graphene-SiC heterostructure symmetrized according to the symmetries in Fig. 3b. The $\vec{K}_{m1}$ and $\vec{K}_{m2}$ vectors fit the positions of hot spots in FS map very well.

Furthermore, the wave vector from the Dirac cone replica at $K_{m1}$ to that at $K_{m2}$ points is
$$\vec{L}_1 = \vec{K}_{m2} - \vec{K}_{m1} = \vec{K}_{Gra}(1,1)$$
as seen from Fig. 3a. This is fully consistent with the experimental results as seen from Fig. 2a and 3b.

Finally, for the constant energy plot away from Dirac point, the triangle-like shapes at the $K_{Gra}(1,1)$ and $K_{Gra}(2,2)$ are opposite to each other (Fig. 3a). Therefore, the Dirac cone replicas on the same side along the $\vec{K}_{Gra}(1,0)$ (corresponding to $-\vec{K}_{m1}$ and $\vec{K}_{m2}$) have the same orientation, but are opposite to the one at $K_{Gra}(1,0)$ (Fig.

3b). This explains the experimental results shown in Fig. 2g.

By following the same way, we can attribute hot spots shown in Fig. 2d as the Dirac cone replicas for moiré structures corresponding to $\vec{G}_{m3}$ and $\vec{G}_{m4}$ in Fig. 3c, with the starting and ending points summarized in Table 1. The locations of these observed Dirac cone replicas fully agree with the corresponding $\vec{K}_{m3}$ and $\vec{K}_{m4}$ vectors in Fig. 3d.

As shown in Fig. 3e-f, the six replicas around the $K_{Gra}$ point, which have more spectra weight with higher energy photons as seen from Supplementary Materials, can also be explained by the moiré structures corresponding to $\vec{G}_{m5}$-$\vec{G}_{m10}$ in Fig. 3e. As seen from Fig. 3e-f, the wave vectors between the six Dirac cone replicas and the nearby main Dirac cone are fixed as

$$\vec{D} = \vec{G}_{Gra}(2,1) - \vec{G}_{SiC}(2,0).$$

This relationship forces these Dirac cone replicas sitting along the high symmetry lines of graphene (Γ-K or K-M directions) and having the same relative distance from the main Dirac cone, as seen from the left K point in Fig. 3f.

We also notice that the wave vector $\vec{L}_2$ between the hot spots at $K_{m6}$ and $K_{m4}$ points (Fig. 3c) is

$$\vec{L}_2 = \vec{K}_{m6} - \vec{K}_{m4} = \vec{G}_{SiC}(0,1),$$

which is consistent with the ARPES data shown in Fig. 3d.

Our results are summarized in Fig.4. We experimental resolve multiply Dirac cone cloning in graphene-SiC heterostructures. The replicas' positions are in good agreements with the conners of moiré Brillouin zones by the graphene-SiC heterostructure, $\vec{K}_{m1}$-$\vec{K}_{m10}$ as summarized in Table 1. The wave vectors connecting Dirac cone replicas at $\vec{K}_{m1}$ and $\vec{K}_{m2}$ is fix as $\vec{K}_{Gra}(1,1)$, and that between replicas at $\vec{K}_{m6} - \vec{K}_{m4}$ is $\vec{G}_{SiC}(1,0)$, further demonstrating the moiré potential origin of the replica bands. The moiré structure scenario is also consistent with the anisotropic shapes of the replica cones. Our results provide a straightforward example of moiré potential in engineering electronic structure with Dirac fermions.

# Figures

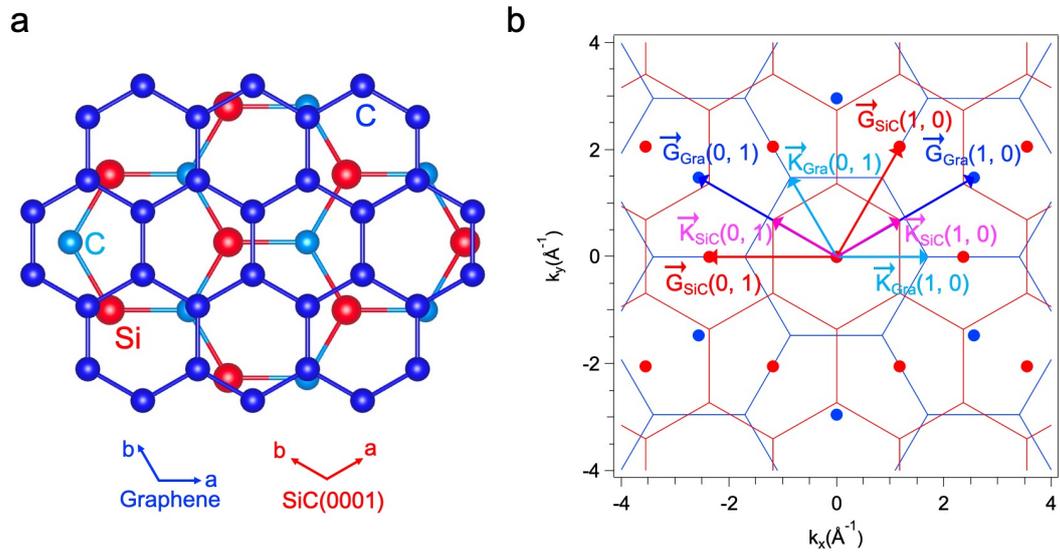

**Figure 1.** Crystal structure (**a**) and Brillouin zones (**b**) of graphene-SiC heterostructure. The primitive vectors of graphene/SiC in real space and reciprocal space are also labeled.

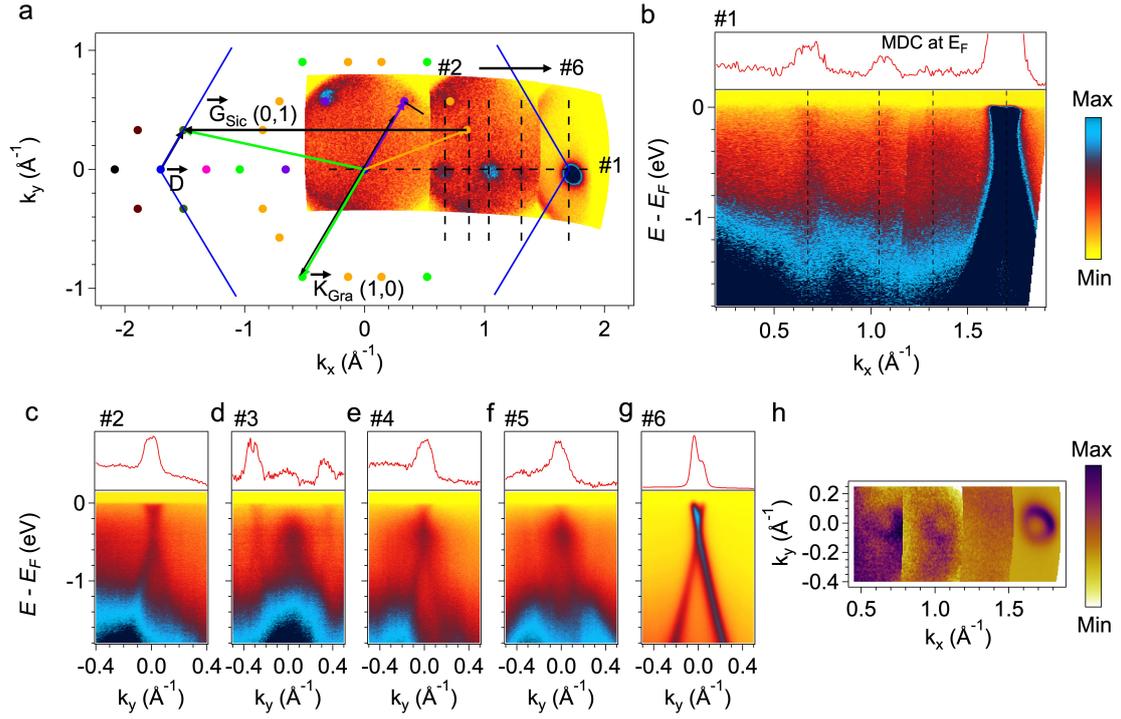

**Figure 2. (a)** Fermi surface mapping of graphene-SiC heterostructure. All the expected hot spots are labeled by colored dots based on the experimentally observed ones and symmetries. **(b)-(g)** Photoemission intensity plots along the cut #1- cut #6, as indicated in **a**. **(h)** Constant energy map at 1 eV below $E_F$.

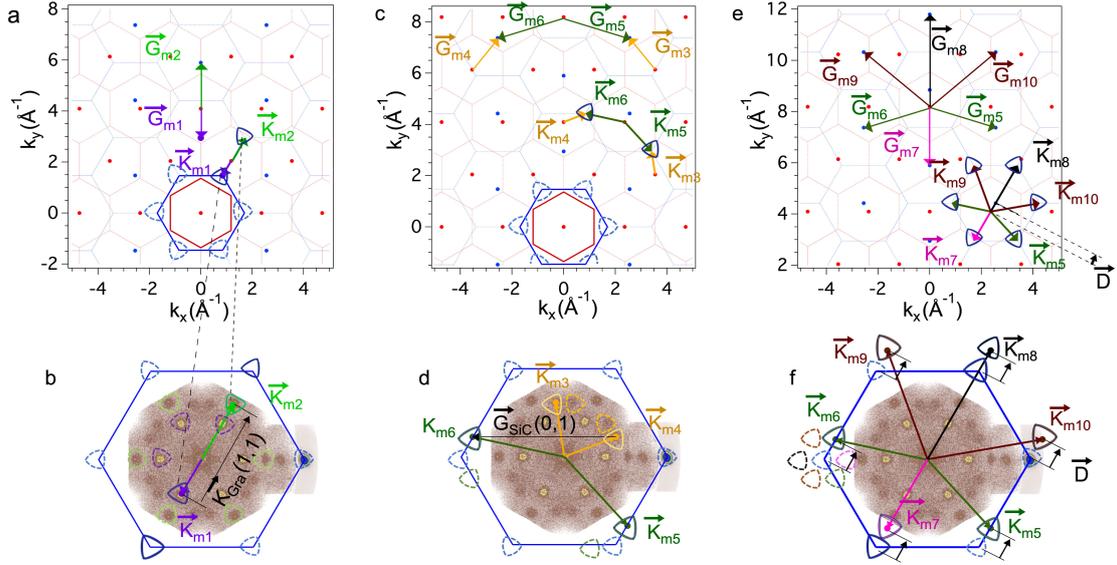

**Figure 3. (a)** The illustration of primitive vectors of moiré 1 and moiré 2 ($\vec{G}_{m1}$ and $\vec{G}_{m2}$) and corresponding $\vec{K}_{m1}$ and $\vec{K}_{m2}$ in reciprocal spaces. **(b)** The direct comparison of $\vec{K}_{m1}/\vec{K}_{m2}$ and experimentally determined Dirac cone replicas. **(c,d)** Same as **(a,b)** but for $\vec{G}_{m3}$-$\vec{G}_{m6}$ and $\vec{K}_{m3}$-$\vec{K}_{m6}$. **(e,f)** Same as **(a,b)** but for $\vec{G}_{m5}$-$\vec{G}_{m10}$ and $\vec{K}_{m5}$-$\vec{K}_{m10}$.

**Figure 4.** Schematic drawing of cloning of Dirac cones by graphene-SiC moiré potential. The corresponding $\vec{K}_m$ vectors are labeled. Some characterized relative distances between the Dirac cone replicas are also indicated.

# Tables

| $n$ | $\vec{G}_{m_n}$ start | $\vec{G}_{m_n}$ end | $\vec{K}_{m_n}$ start | $\vec{K}_{m_n}$ end |
|---|---|---|---|---|
| 1 | $\vec{G}_{SiC}(2,1)$ | $\vec{G}_{Gra}(1,1)$ | $\vec{K}_{SiC}(2,1)$ $(\vec{G}_{SiC}(1,0))$ | $\vec{K}_{Gra}(1,1)$ |
| 2 | $\vec{G}_{SiC}(2,1)$ | $\vec{G}_{Gra}(2,2)$ | $\vec{K}_{SiC}(2,1)$ $(\vec{G}_{SiC}(1,0))$ | $\vec{K}_{Gra}(2,2)$ |
| 3 | $\vec{G}_{SiC}(3,0)$ | $\vec{G}_{Gra}(3,2)$ | $\vec{K}_{SiC}(3,0)$ | $\vec{K}_{Gra}(3,2)$ |
| 4 | $\vec{G}_{SiC}(3,3)$ | $\vec{G}_{Gra}(2,3)$ | $\vec{K}_{SiC}(3,3)$ | $\vec{K}_{Gra}(2,3)$ |
| 5 | $\vec{G}_{SiC}(4,2)$ | $\vec{G}_{Gra}(3,2)$ | $\vec{K}_{SiC}(4,2)$ $(\vec{G}_{SiC}(2,0))$ | $\vec{K}_{Gra}(3,2)$ |
| 6 | $\vec{G}_{SiC}(4,2)$ | $\vec{G}_{Gra}(2,3)$ | $\vec{K}_{SiC}(4,2)$ $(\vec{G}_{SiC}(2,0))$ | $\vec{K}_{Gra}(2,3)$ |
| 7 | $\vec{G}_{SiC}(4,2)$ | $\vec{G}_{Gra}(2,2)$ | $\vec{K}_{SiC}(4,2)$ $(\vec{G}_{SiC}(2,0))$ | $\vec{K}_{Gra}(2,2)$ |
| 8 | $\vec{G}_{SiC}(4,2)$ | $\vec{G}_{Gra}(4,4)$ | $\vec{K}_{SiC}(4,2)$ $(\vec{G}_{SiC}(2,0))$ | $\vec{K}_{Gra}(4,4)$ |
| 9 | $\vec{G}_{SiC}(4,2)$ | $\vec{G}_{Gra}(3,4)$ | $\vec{K}_{SiC}(4,2)$ $(\vec{G}_{SiC}(2,0))$ | $\vec{K}_{Gra}(3,4)$ |
| 10 | $\vec{G}_{SiC}(4,2)$ | $\vec{G}_{Gra}(4,3)$ | $\vec{K}_{SiC}(4,2)$ $(\vec{G}_{SiC}(2,0))$ | $\vec{K}_{Gra}(4,3)$ |

**Table 1.** The starting and ending points of $\vec{G}_{m1}$-$\vec{G}_{m10}$ and $\vec{K}_{m1}$-$\vec{K}_{m10}$.